# Interpreting X-ray absorption spectra of Vanadyl Phthalocyanines Spin Qubit Candidates using a Machine Learning-Assisted Approach

J.H. Lee[1,2], C. Urdaniz[1,3], S. Reale[1,3,4], K.J. Noh[1,2], D. Krylov[1,3], A. Doll[5], L. Colazzo[1,3], Y.J. Bae[1,2], C. Wolf[1,3] and F. Donati[1,2]

[1] Center for Quantum Nanoscience (QNS), Institute for Basic Science (IBS), Seoul 03760, Republic of Korea

[2] Department of Physics, Ewha Womans University, Seoul 03760, Republic of Korea

[3] Ewha Womans University, Seoul 03760, Republic of Korea

[4] Department of Energy, Politecnico di Milano, Milano 20133, Italy

[5] Swiss Light Source (SLS), Paul Scherrer Institut (PSI), 5232 Villigen, Switzerland

**ABSTRACT**

The magnetic dilution of Vanadyl phthalocyanine (VOPc) within the isostructural diamagnetic Titanyl phthalocyanine (TiOPc) affords promising molecular spin qubit platforms for solid-state quantum computing. The development of quantitative methods for determining how the interactions with a supporting substrate impact the electronic structure of the system are fundamental to determine their potential integration in physical devices. In this work we propose a combined approach based on X-ray absorption spectroscopy (XAS), atomic multiplet calculations, and density functional theory (DFT) to investigate the 3$d$ orbital level structure of VOPc on TiOPc/Ag(100). We characterize VOPc in different molecular environments realized by changing the thickness of TiOPc interlayer and adsorption configuration on Ag(100). Depending on the molecular film structure, we find characteristic XAS features that we analyze using atomic multiplet calculations. We use a Bayesian optimization algorithm to accelerate the parameter search process in the multiplet calculations and identify the ground state properties, such as the 3$d$ orbital occupancy and splitting, as well as intra-atomic interactions. Our analysis indicates that VOPc retains its spin S = ½ character in all configurations. Conversely, the energy separation and sequence of the unoccupied V 3$d$ orbitals sensitively depend on the interaction with the surface and TiOPc interlayer. We validate the atomic orbital picture obtained from the multiplet model by comparison with DFT, which further allows us to understand the VOPc electronic properties using a molecular orbital description.

## I. INTRODUCTION

Magnetic molecules with electron spin S=1/2 can serve as a native platform for the realization of molecular qubits for quantum logic devices[1]. Extending the coherence time of quantum systems enables more robust and reliable quantum computations. The rational design of molecular building blocks can be used to fine-tune the spin environment and coherence properties, e.g., spin-phonon coupling or noises occurring from the fluctuations of neighboring nuclear and electron spins[2-5]. Integrating molecular spin qubits with solid-state surfaces introduces additional sources of decoherence.

To this extent, it is crucial to identify suitable molecules/substrate combinations and characterize the electronic and magnetic configuration of molecular spin qubits upon surface adsorption [6, 7].

Metal phthalocyanines (MPcs) are a class of molecular compounds that has been extensively investigated for potential applications as single molecule qubits. Their chemical properties and structural robustness as well as the possibility to chemically modulate the hybridization between the metal center and the molecular orbitals lead to well localized spins[8, 9], and in turn long spin lifetime and coherence time[10]. Among the various MPc variations, Vanadyl Phthalocyanine (VOPc) has a robust $3d^1$ configuration, and a long coherence time of up to 1 $\mu s$ at room temperature[11], making it a promising molecular spin qubit[8, 12-15]. Although the V center typically retains the pristine spin=1/2 when adsorbed on metal and semiconducting surfaces[13], the ligand may be strongly affected due to the hybridization with surface electrons. Recent studies of VOPc molecules on top of graphene/SiC(0001)[16] and TiOPc/Ag(100)[17] indicate weak coupling with the surface, which preserve the original ligand structure. However, fundamental characteristics such as the crystal field of such surface-supported molecular spins system and their potential as fundamental unit in the creation of 2D nanostructures remain largely unexplored.

Synchrotron techniques such as x-ray absorption spectroscopy (XAS), circular dichroism (XMCD), and linear dichroism (XLD) have been widely used to characterize molecular spin qubits[15-17]. Several approaches have been explored to interpret X-ray spectra of metal-organic complexes[18]. Although ab-initio approaches can be used to this extent, the strong impact of the inter-atomic core-hole interaction is typically not well captured by these methods and ad-hoc adaptations are often required[19-21]. Conversely, multiplet calculations offer a viable way to simulate spectra corresponding to the excitation of $2p$ core electrons of transition metals ions into the valence $3d$ and $4s$ states[22] and retrieve the magnetic properties of the investigated systems. A direct comparison with experiments can validate the inferred electronic configuration and crystal field structure. However, due to the challenges with fitting data using a large number of free parameters, it has been so-far accepted to restrict the number of parameters and value ranges to conduct a qualitative comparison between experiment and theory[23]. In the recent years, the field has been evolving towards more quantitative approaches that aim at supplying the required input parameters using DFT[24] or by tackling the issue by advanced computational methods such as adaptive grid algorithm[25]. Notwithstanding, due to the presence of sharp features in the XAS and XMCD spectra, quantitative modeling of the VOPc data with multiplet calculations remained challenging [8, 9, 15, 16].

To this end, machine learning-based optimization algorithms can efficiently be used to analyze systems with multiple control parameters[26]. Among the different methods, Bayesian optimization (BO) is one of the most utilized model-based approach for globally optimizing expensive black-box functions with high correlation between parameters[27, 28]. This approach offers two main advantages, namely, sampling efficiency and robustness under noisy observations[29]. Combining BO with multiplet calculations is expected to be effective in optimizing X-ray spectra fitting and infer the orbital level structure. Several authors have explored the use of machine learning techniques to interpret X-ray absorption spectra [30, 31]. Although machine learning methods are being increasingly applied to the interpretation of $K$-edge XAS, i.e. the excitation of a core electron from $1s$ states [32-35], less attention has been given to XAS at the *L*-edge, i.e., from the 2*p* states. Due to the spin-orbit splitting of the 2*p* states, the physical processes at the *L*-edge are generally more complex than at the *K*-edge. XAS and XMCD at the *L*-edge, however, offer the advantage of carrying information on the electron spin of the valence states, and permits the investigation of the magnetic properties of materials[36].

Here we report a machine learning-based approach for the interpretation of the XAS and XMCD measurements of VOPc adsorbed on Ag(100) and on TiOPc/Ag(100). Combining multiplet calculations with BO we quantitatively reproduce the spectral features of VOPc at the V $L_{2,3}$ edges. This approach provides a substantial speedup with respect to conventional fitting procedures[25] and allows us to infer the electronic configuration and 3*d* orbital level structure for several molecular film configurations and TiOPc interlayers. Our results indicate that the pristine orbital level scheme of V within VOPc can be modified by the interaction with the surface and the TiOPc interlayer. Comparison with results obtained from density functional theory (DFT) calculations further allows us to interpret the outcome of our approach and obtain insight into the orbital structure reconfiguration induced by the different adsorption geometries.

The paper is organized as follows. In Sec. II we provide details for the experimental and theoretical methods, including the sample preparation, the XAS measurements, the sum rules analysis, the multiplet calculations, and Bayesian optimization procedure used to fit the data. Details about the DFT are also included in this section. Section III presents the results obtained by XAS, multiplet calculations, and DFT that reveal the 3*d* orbital splitting and spin configuration of VOPc on both TiOPc/Ag (100) and bare Ag(100). Sec. IV summarizes the main conclusions of our work.

**II. Details of the experiment and modelling**

    **A. Sample Preparation**

The Ag(100) surface was prepared in ultra-high vacuum (UHV) by repeated cycles of sputtering at Ar$^+$ pressure of ~5.0 x 10$^{-6}$ Torr and annealing at 700K until obtaining atomically flat terraces, as verified prior to the X-ray experiments using a Omicron variable temperature scanning tunneling microscope (STM) available at the beamline. TiOPc (99% purity) and VOPc (99% purity) were purchased from Sigma-Aldrich and used without further purification. Before deposition, the molecules were thermally outgassed at 620 K for several hours in the UHV chamber. TiOPc was sublimated at 609 K and VOPc was sublimated at 593 K using a commercial evaporator from Kentax. The Ag substrate was kept at room temperature during the molecule deposition to avoid intermixing of the two species [17]. The TiOPc and VOPc coverage was estimated using STM and correlated with the amplitude of the XAS at the Ti and V $L_{2,3}$ edges, respectively. Two sets of VOPc on TiOPc/Ag(100) heterostructures were made by depositing 0.5 monolayers (ML) VOPc on (1.1 ML TiOPc)/Ag and 1.0 ML VOPc on (2.0 ML TiOPc)/Ag. The sample with 0.7 ML of VOPc on Ag(100) surface was characterized with X-rays before and after annealing it at 513 K. The samples were transferred to the X-ray measurement stage without breaking the vacuum.

    **B. X-ray Absorption Spectroscopy**

XAS, XMCD and XLD measurements of VOPc on TiOPc/Ag(100) were performed at the EPFL-PSI X-treme beamline[37]. The samples were prepared *in situ* with the method explained above. The measurements were performed at low temperature (2.5 K), and at magnetic fields of up to 6.8 T. The X-ray beam from the synchrotron source was aimed at the sample, either at 0° (normal incidence, NI) or 60° angle (grazing incidence, GI) with respect to the surface normal, with the photon beam parallel to the magnetic field. These angle-dependent measurements allow exploring the anisotropy of the charge distribution and of the magnetic moments [38]. The choice of the grazing incidence (GI) angle was determined to have the largest angle of incidence with respect to the surface normal that allows the X-ray beam to fully hit the sample surface without extending beyond its rim. In our experiment, we used a defocused photon beam spot (approximately 1 mm2 in size) to minimize the photon flux and avoid beam-induced degradation of the molecular layer. The crystals used in our experiments (5 mm diameter) offer an elliptical cross section of 2.5x5 mm when turned to 60 degrees from the normal, which is safely larger than the photon beam spot size and allow us to avoid probing the less homogenous surface regions close to the sample rim. The circular polarizations (CR and CL) are defined based on the direction of the photon beam, while the linear polarizations (LH and LV) are defined based on the sample surface. All data were collected in total electron yield mode and normalized to the related pre-edge intensity. For both circular and linear polarization measurements, XAS is obtained as the sum of the two polarizations, XMCD is the difference between CR and CL, while XLD is obtained as the difference between LH and LV. Background data was collected on a clean Ag (100) and subtracted from the spectra of samples with molecular films. The XAS units in all plots show the relative change in absorption from the pre-edge.

### C. Sum rules

The difference in absorption between left and right-circularly polarized X-rays, normalized over the $L_3$ and $L_2$ edges, can be expressed as a function of the XAS and XMCD integrals yielding the expectation value of the effective spin $m_{s,eff}$[36] and orbital moment $m_l$[39] projected along the photon beam axis. These so-called sum rules are applicable when transitioning between two distinct shells, such as exciting from a 2*p* core state to 3*d* valence states in transition metal systems, where these 3*d* valence states are assumed to be distinct from other final states. The total magnetic moment per atom can be represented as follows:

$$m_{s,eff} = m_s + m_D^\alpha = -\frac{\mu_B}{\hbar}(2\langle S_z\rangle + 7\langle T_z\rangle) = -n_h\frac{6p-4q}{t}, \quad \text{(Eq. 1)}$$

$$m_l = -\frac{\mu_B}{\hbar}\langle L_z\rangle = -n_h\frac{4}{3}\frac{q}{t}, \quad \text{(Eq. 2)}$$

where $m_D^\alpha$ is angle dependent intra-atomic dipole moment, $\mu_B$ is the Bohr magneton, $n_h$ is number of holes, and the expectation values of the orbital and spin magnetic moments of the valence electrons, denoted as $\langle L_z\rangle$ and $\langle S_z\rangle$ respectively. These calculations are based on the values of *p* and *q* that correspond to the integral of XMCD solely over the $L_3$ and over both the $L_2$ and $L_3$ edges, respectively while *t* represents the integral area of XAS. The value of the spin from sum rules is mixed with the expectation value of the intra-atomic magnetic dipole, $\langle T_z\rangle$. To obtain the actual spin angular momentum, we need to compute the value given by Eq. 3 by removing $\langle T_z\rangle$ as follows:

$$\langle S_z\rangle = -\frac{1}{2}\left(\frac{\hbar}{\mu_B}m_{s,eff} + 7\langle T_z\rangle\right). \quad \text{(Eq. 3)}$$

For VOPc, the $\langle T_z \rangle$ value, can be determined from the literature assuming the electron is occupying an atomic-like $3d_{xy}$[9] orbital giving $\langle T_z \rangle = -0.28\ \hbar$[38].

### D. Bayesian-optimization assisted multiplet calculations

To analyze the acquired data, we compared the experimental results with calculated X-ray spectra at the V $L_{2,3}$ edges using the Quanty multiplet code[40], which includes the electron-electron interaction, spin-orbit coupling, and the splitting of 3*d* orbitals induced by the crystal field. The values for the Slater integrals and spin-orbit coupling were determined through atomic calculations with the Cowan code[41]. The spectra were calculated using the imaginary part of the Green's functions of the electric dipole transition operator and the V electron wavefunctions[40], with a Lorentzian broadening of 70 meV to match the linewidth of the lowest energy X-ray transitions. A linear photon energy-dependent Gaussian broadening was also applied from 0.036 to 0.89 to reproduce the energy-dependent lifetime of the X-ray absorption excitation[42]. The free parameters of the model are the rescaling factors for the 3*d*-3*d* and 2*p*-3*d* atomic Slater integrals (2 parameters) and the relative on-site energy of the 3*d* orbitals (3 parameters). An additional scaling parameter is used to match the overall amplitude of the simulation with the experiment. In the calculations we assumed the $b_{2g}$ orbital as the lowest lying-in energy[9], while the splitting of the others was included considering the $C_{4v}$ symmetry of the V $b_{1g}$, $a_{1g}$, $e_g$ orbitals.

The values of these parameters have been determined fitting the 6 independent absorption spectra, i.e. circular XAS/XMCD in NI and GI at 6.8T, and linear XAS/XLD in GI at 0.05T, with a BO algorithm implemented in MATLAB. In a typical BO process, the purpose is to minimize an objective function [43]. To do so, at each iteration, a prior distribution is used as the input for an acquisition function that identifies the coordinates on the parameter space on which to pick the next data point. After observing the objective function at the picked parameter set, we fit the sampled data with a Gaussian process regression as implemented in the MATLAB function *fitgrp*, obtaining a posterior distribution characterized by a mean and a standard deviation. The posterior is then used as a prior for the next iteration. In our case, the acquisition function is a 2-sigma lower confidence bound. No input is given to choose the initial prior. Instead, we indicate the number of points that are sampled randomly over the parameter space before the regression (1 point in our case). Using these points, a Gaussian process regression is used to obtain an initial posterior distribution. This posterior is then used as prior to the following iteration. A partial dependence representation of the final posterior obtained for the 4 samples investigated in this work is given in appendix F. As for the objective function, we use an error function containing the sum of two terms:

$$Err\ =\ \sum_\alpha \frac{RSS_\alpha}{TSS_\alpha} + W \sum_\alpha \sum_i \left(P_{\alpha,i} - \hat{P}_{\alpha,i}\right)^2. \quad \text{(Eq. 4)}$$

In Eq. 4, the first term is ratio between the residual sum of squares $RSS_\alpha$ computed for each experimental/simulated spectra pair and the total sum of squares ($TSS_\alpha$) of the experimental spectra, with $\alpha$ labeling each of the 6 XAS, XMCD, and XLD. To calculate $RSS_\alpha$, the simulated data were interpolated to match the experimental energy axis. The second term is the discrete set of single points which is computed as the difference between experimental ($P_{\alpha,i}$) and calculated ($\hat{P}_{\alpha,i}$) intensity of the most relevant features. The subscript "i" indicates the individual feature. This additional term in the objective function does not include a contribution from the XAS and only two features per XMCD and XLD are considered for a total of 6 contributions for each sample. These features are marked as gray dashed lines in each X-ray spectra shown in the text. Through the first term, the fitting procedure tends to prioritize matching broader and smoother features, which typically shows little variations with the molecular structure. Conversely, the second term is required to give weight to the sharp and intense features that are more sensitive to changes in the local environment and are crucial to capture the underlying physics (see Appendix D), hence they represent an additional figure of merit of the fit. This strategy was also discussed in a previous work, where the use of descriptors (such as amplitudes, slopes, and curvatures at specific spectral features) allows enhancing the match between the fit and the spectra[44]. For vanadium in VOPc showing sharp spectroscopic features, we found that using the amplitude as the sole descriptor of the system enabled us to accurately track the variations of the spectra as a function of the molecular film. Equivalently accurate fitting results can also be obtained by selecting an energy windows around the features of interest (see Appendix E). For the second terms, we chose a value of the coefficient $W$ considering the weight between the number of points included in the $P_{\alpha,i}$ terms (up to 6) and the number of points in each spectrum (400) used to calculate the RSS/TSS term. As the ratio between the number of points between these two terms is of the order of hundreds, the RSS/TSS terms are about hundred times larger than $P_{\alpha,i}$ terms. Therefore, assigning W of the order of hundreds allows us to balance the weight of the two terms and to give them comparable relevance in the optimization procedure. The specific value of W = 300 was chosen after testing values from 100 to 500, and noticing that accurate fitting of the sharp features could only be achieved for W ≥ 300, as this value gives enough weight to the $P_{\alpha,i}$ terms.

The flow-chart of the combined BO-multiplet process is outlined in Fig. 1. We initiated the optimization with random parameters and performed multiplet calculations. The spectra obtained from the calculations were then compared with the experimental data, to compute the error. We use BO to minimize the error function. At each iteration, the error value calculated for the set of parameters is used as observation point to update the objective function of the BO. Due to the large number of local minima in the error function, it is not straightforward to identify terminating conditions for the fitting procedure to pinpoint a global minimum. In this work, we chose to overcome this issue by terminating the procedure after a fixed number of steps (300) and repeating the same procedure several times (30 trials), selecting the best 6 outcomes providing the lowest value of the error function. This approach allows us to 1) avoid the characteristic slowing down of the BO fitting with increasing number of steps, 2) explore potential solutions comprehensively, and 3) discarding solutions that remained pinned to a local minima. The simulated results presented in this paper are shown as the average and standard deviation of the the best 6 runs.

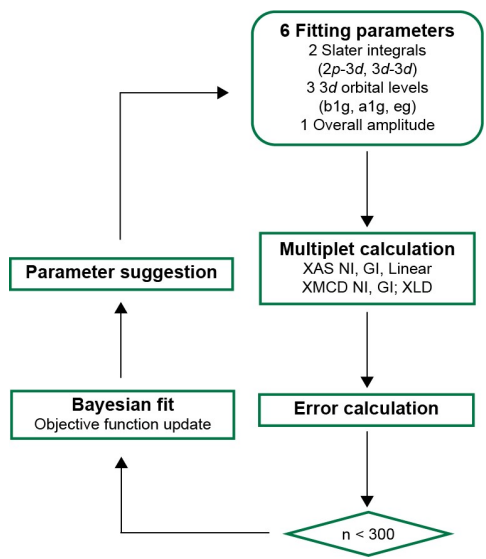

*Figure 1 Flow-chart of the Bayesian optimization combined with multiplet calculations used to fit the experimental XAS, XMCD, and XLD spectra. The termination condition is based on the number of runs (n), as described in the text.*

### E. Density Functional Theory

We calculated the electronic structure for all adsorption configurations using DFT as implemented in Quantum Espresso (versions 6.8 and 7.0)[45]. All ions were represented using projector augmented-wave pseudopotentials from the PSLibrary[46] and the PBE functional was employed to approximate the exchange-correlation functional[47]. Dispersive forces were treated using the revised VV10 method (rvv10). We used rVV10 throughout this work as it is the most reliable dispersion correction for systems with mixed bonding environments[48]. The cells were modeled using suitable lateral supercells of the relaxed simple unit cells, padded with approximately 2.0 nm of vacuum in the *z*-direction, and decoupled from the periodic images in the *z*-direction using dipole correction. Energy and charge density cutoffs were set at 60 Ry and 520 Ry, respectively. Only the Gamma point was used for integration of the Brillouin zone. A Hubbard *U* correction of 3.6 eV and 2.1 eV was applied to the 3*d* electrons of the Ti and V atoms, respectively. We implemented a Hubbard U correction on the transition metal core of the molecules to correct the position of localized 3d states. A U value of 3.6 eV for Ti(3d) and 2.1 for V(3d) successfully reproduces the values and orbital character of the gap, consistent with what has been previously reported[17, 49].

## III. Results and discussion

### A. VOPc on 2ML TiOPc/Ag (100)

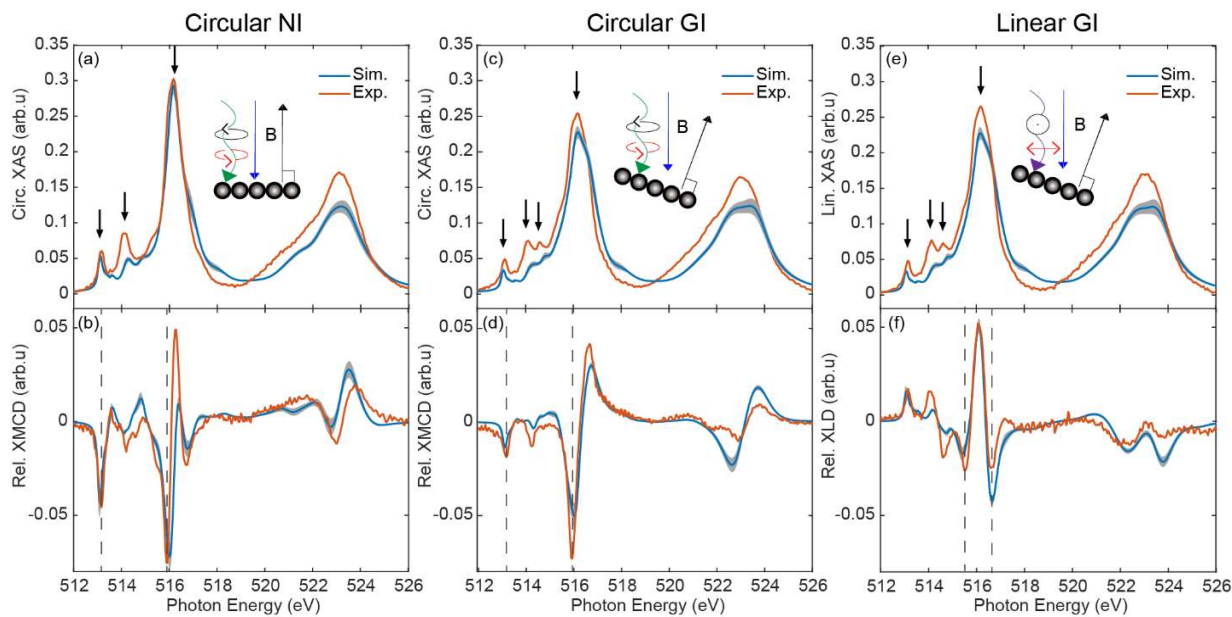

*Figure 2: Experimental and simulated XAS, XMCD, and XLD spectra for VOPc on 2 ML of TiOPc/Ag (100). (a) XAS and (b) XMCD spectra at 0 ° normal incidence (NI) at B = 6.8 T, (c) XAS and (d) XMCD spectra at 60 ° grazing incidence (GI) at B = 6.8 T, (e) XAS with linear polarization and (f) XLD spectra at GI at B = 0.05 T. The red lines show the experimental data, while blue lines represent the average of the best 6 simulations. The shaded gray area indicates the standard deviation from the mean for the simulations. The green arrow indicates left and right circularly polarized light, and the purple arrow indicates vertical and horizontal light. Peaks marked with black arrows in panels (a), (c), and (e) indicate single electron-like transitions from $2p_{3/2}$ core levels to unoccupied 3d orbitals, as discussed in the text. The features considered to compute the second term in Eq. 4 are indicated by gray dashed lines in (b), (d), and (f).*

Figure 2 shows the total XAS, XMCD, and XLD of VOPc deposited on 2 ML of TiOPc/Ag(100). This sample was analyzed at different incidence angles (NI, for Figs. 2a-b, and GI, for Figs. 2c-f) obtained from the sample for which the VOPc is mostly separated from the Ag(100) substrate by TiOPc, i.e., VOPc deposited on 2 ML of TiOPc/Ag(100). The spectra show a strong resemblance to those reported in previous studies of VOPc on graphene/SiC(0001)[16]. The XAS in NI is characteristic of the $3d^1$ configuration and shows three distinct excitations at the $L_3$ edge (see arrows in Fig. 2a), whereas the GI XAS shows an additional peak in the spectrum (see arrows in Fig. 2c, 2e). These features have been previously assigned to single electron-like transitions from $2p_{3/2}$ core levels to unoccupied $3d$ orbitals based on their angle-dependent intensity [9]. For systems with pronounced anisotropic charge distribution such as metal-organic complexes, the XAS intensity varies with the beam's incident angle relative to the axis of the molecules, particularly when they are well-oriented on the surface [9], [50]. Therefore, also for the present system the variations XAS intensity with respect to the angle supports the observation of a flat adsorption on the surface [17].

The $L_2$ peak is visible as a single with the maximum located at 523.8 eV. This peak is composed of several excitations from the $2p_{1/2}$ characterized by a broader linewidth compared to the one of $L_3$ edge transitions. This additional broadening stems from shorter lifetime of the excited states caused by the $L_2$-$L_3$ Coster-Kronig decay[42].

Figure 2b and d show the XMCD obtained for NI and GI, respectively. In all XMCD spectra, a prominent feature appears at 515.9 eV close to the maximum of the $L_3$ XAS peak while less intense features are observed at lower energies and at the $L_2$ edge. The large XMCD signal indicates a localization of the magnetic moments in the 3*d* orbitals. Comparison with previous studies confirms that the shape of the XMCD spectrum is characteristic of a *S*=1/2 system[9, 15, 16]. Finally, the pronounced XLD observed in Fig. 2f indicates that the VOPc molecules adsorb with the molecule plane parallel to the the underlying substrate surface[9, 15, 16, 50].

Applying the sum rules (Eqs. 1–3) to the spectra obtained at NI, we find $\langle L_z \rangle = 0.05 \pm 0.01\ \hbar$ and $\langle S_z \rangle = 0.32 \pm 0.07\ \hbar$. The presence of a small but non-vanishing orbital angular momentum is in line with a previous analysis[51]. However, the value of the spin from the sum rules is significantly lower than what expected for a *S* = ½ configuration. In order to gain a better understanding of the electronic configuration at the V center, we utilized the BO to fit the experimental results with the spectra obtained from multiplet calculations[40]. The simulated spectra obtained through multiplet calculations nicely fit the experimental data (see Fig. 2), with the corresponding parameters listed in Tab. 1 and Tab. 3 in the Appendix. The position in energy of all the relevant spectral features in simulated XAS, XMCD, and XLD are in excellent agreement with the experiment. In addition, the intensities are also well matched at the $L_3$ edge, while minor discrepancies are visible at the $L_2$ edge. We ascribe the latter to the additional weight given to the corresponding $L_3$ XMCD features in the fitting procedures, as described in the methods section.

The V atom in the VOPc ligand has a five-coordinated square-pyramidal geometry and is shifted above the plane of the four neighboring nitrogen atoms[52]. In this geometry, energy splitting of the V orbitals can be rationalized using the $C_{4v}$ symmetry point group. The related splitting and occupation of the 3*d* orbitals obtained from the multiplet calculations is shown in Fig. 3. In accordance with previous studies[9, 13-17], our findings indicate that the V center is in a tetravalent state ($V^{4+}$), with a single unpaired electron located on the $d_{xy}$ orbital. The expectation value of $\langle L_z \rangle_{\text{multiplet}} = -0.01\hbar$ obtained from multiplet calculations agrees well in magnitude with the sum rule value, while $\langle S_z \rangle_{\text{multiplet}} = 0.49\hbar$ largely approaches the theoretical value for the *S*=1/2 system. This result indicates that the spin sum rule applied to the experimental spectra underestimates the spin value of V possibly due to the $L_3$-$L_2$ edges overlap affecting the calculation of the integrals over the XMCD[53].

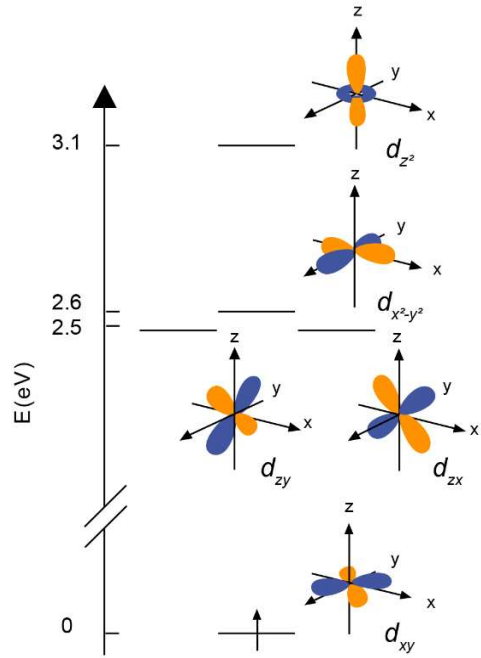

*Figure 3 Schematic diagram showing the electronic configuration of V and the splitting of its 3d-orbitals. For VOPc on 2ML of TiOPc/Ag (100) as obtained from the optimization of the crystal field from the multiplet calculations*

Figure 3 shows the first unoccupied 3$d$ orbitals ($d_{zx/zy}$) separated from the ground state by 2.48 ± 0.08 eV, followed by the $d_{x^2-y^2}$ (2.59 ± 0.10 eV) and by the $d_{z^2}$ (3.13 ± 0.05 eV). The energy range of the orbital splitting is consistent with previous literature[51]. Interestingly, our analysis indicates that the sequence between the $d_{x^2-y^2}$ and $d_{z^2}$ orbitals is reversed with respect to that proposed in a previous study[9, 51]. Due to the orbital splitting provided by ligand, the relative energy of $d_{z^2}$ with respect to the $d_{x^2-y^2}$ is ultimately determined by the shift of the V atom with respect to the plane of the four neighboring nitrogen atoms and by its proximity to the O atom[54]. The effect of the adsorption configurations on different substrates[12] or different molecular films[17] may lead to a distortion of the pristine molecular structure and, consequently, to a different sequence of the orbital levels. This mechanism could explain the differences between our results and previous studies[9, 51], as it will be further elaborated in the following sections.

### B. Influence of interlayer thickness and molecule orientation on the electronic properties of VOPc

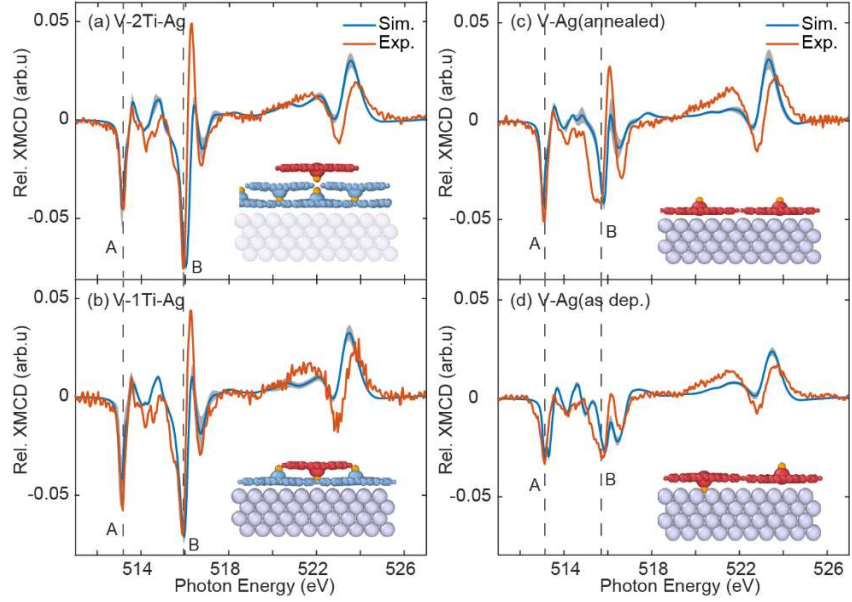

*Figure 4 Experimental and simulated XMCD NI spectra for VOPc are given as a function of interlayer thickness: (a) VOPc on 2ML of TiOPc on Ag (100), V-2Ti-Ag (same as Fig.2b), (b) VOPc on 1ML of TiOPc on Ag (100), V-1Ti-Ag, (c) VOPc on Ag(100) after annealing at 570K, with VOPc showing oxygen up configuration, V-Ag(annealed), and (d) VOPc on Ag(100) as deposited, V-Ag(as dep.). The latter shows 50% oxygen up and 50% oxygen down. The red lines show the experimental data, while blue lines represent the average of the best 6 simulations. The corresponding standard deviation from the mean value at each point is shaded in gray. The features considered to compute the second term in Eq. 4 are indicated by gray dashed lines. The molecular structure obtained from DFT is shown for each different configuration as an inset. For the sample (a), the Ag substrate(shaded) was not included in the DFT calculation.*

To investigate the impact of the TiOPc interlayer thicknesses and the VOPc orientation on the orbital sequence and magnetic properties, we compare the spectra of VOPc on 2 ML TiOPc/Ag(100) (Fig. 4a, V-2Ti-Ag, same as Fig.2b) with different molecular structures: VOPc on 1 ML TiOPc interlayer (Fig. 4b, V-1Ti-Ag), and VOPc directly on Ag(100) surface both as deposited (Fig. 4d, V-Ag(as dep.)) and after annealing at 570 K (Fig. 4c, V-Ag(annealed)). Previous DFT calculations and STM measurements on TiOPc/Ag(100) indicated a 50%-50% oxygen-up to oxygen-down ratio upon surface adsorption on Ag(100), which changed to 100% oxygen up after annealing[17]. We expect VOPc to exhibit a similar behavior, given their similar molecular structure. While the XAS shape was similar in all samples (see Figs. 8-10 in Appendix), we observe significant differences in the XMCD spectra in NI (Fig. 4). Prominent features of the $L_3$ edge at the energy of 513.1 eV and of 515.3 eV - 515.9 eV (indicated as peaks A, B in the Figs. a–d) exhibit variations in their intensities depending on the TiOPc interlayer thickness and VOPc adsorption configuration. In VOPc, the $L_3$ peaks are well described by single electron-like transitions to unoccupied $3d$ orbitals[9]. Therefore, the variation of the intensities observed for the A and B peaks reveal possible modifications of the crystal field depending on the molecular adsorption configuration. In addition, these peaks are quite sharp for samples with TiOPc interlayers, while VOPc/Ag(100) samples show lower intensities and broader peaks. This effect might be due to the stronger hybridization between V states and Ag conduction electrons happening in the absence of TiOPc interlayer. For all samples, the sum rules give $\langle S_z \rangle$ values are lower than what expected for $S = ½$ (see Tab. 2 in the Appendix). As described above, the discrepancy related to $\langle S_z \rangle$ can be ascribed to an underestimation of the spin sum rule due the overlap between the $L_3$ and $L_2$ edges[53].

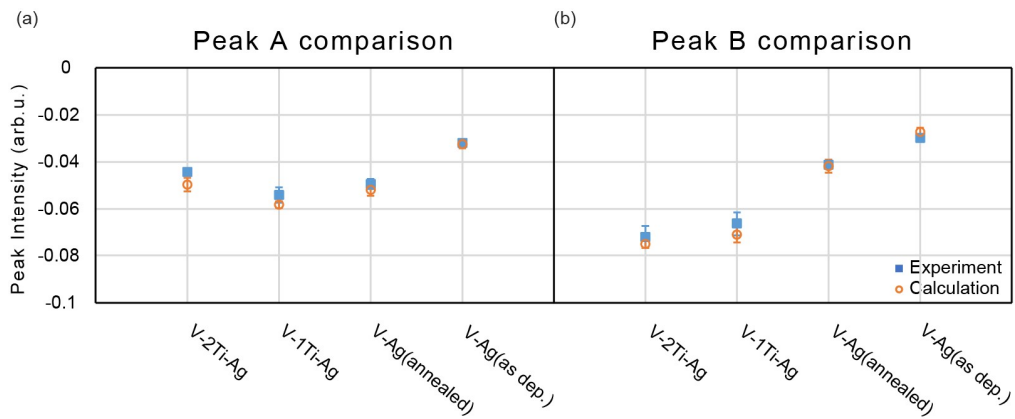

*Figure 5 Comparison between experiment and multiplet calculations for the intensity of the peak A (a) and peak B (b) shown in Fig. 4. For the experiments, we considered the average of the three points at the tip of each peak and the error bars show the standard deviation over these three points. For multiplet calculations, we considered the average of the tipping points of each peak over 6 calculations and the error bars show the standard deviation over these six values.*

|  | V-2Ti-Ag | V-1Ti-Ag | V-O$^{up}$ | V-O$^{dn}$ |
|---|---|---|---|---|
| $A_{1g}$ ($d_{z^2}$) | 3.13 $\pm$ 0.05 eV | 3.94 $\pm$ 0.18 eV | 2.95 $\pm$ 0.02 eV | 2.67 $\pm$ 0.06 eV |
| $B_{1g}$ ($d_{x^2-y^2}$) | 2.59 $\pm$ 0.10 eV | 2.44 $\pm$ 0.16 eV | 2.16 $\pm$ 0.06 eV | 2.67 $\pm$ 0.05 eV |
| $E_g$ ($d_{xz,yz}$) | 2.48 $\pm$ 0.08 eV | 2.35 $\pm$ 0.14 eV | 2.11 $\pm$ 0.06 eV | 2.08 $\pm$ 0.06 eV |
| $B_{2g}$ ($d_{xy}$) | 0 eV | 0 eV | 0 eV | 0 eV |

*Table 1 Crystal field parameters for multiplet calculations obtained from the average of the best 6 fits performed using the BO. The energies are reported as difference from the $B_{2g}$ level.*

To interpret the variations of the spectra with the molecular layer structure, we fit our data with multiplet calculations using the Bayesian optimization (see Fig. 4 and Appendix Figs. 8-10; see Table 1 and Appendix Table 3 for the fit parameters). Our method allows us to follow the trend of the A and B peaks (Fig. 5), providing an accurate modeling of the electronic and magnetic properties of the samples. Our calculations reveal that the V atom retains a S = 1/2 in all four cases, even when the VOPc molecule is directly adsorbed on the Ag surface (Appendix Table 2). In Tab. 1, we summarize the orbital energies obtained from the fit to the experiment, where the orbitals for the VOPc adsorbed with the oxygen up (V-O$^{up}$) are inferred from the annealed VOPc/Ag(100) sample, while for the oxygen down case (V-O$^{dn}$) the orbitals are extracted from the simulated as-deposited VOPc/Ag(100) with an assumed one-to-one ratio between V-O$^{up}$ vs V-O$^{dn}$ (as previously observed for TiOPc on Ag(100)[17]). For the unoccupied states, the separation to the $d_{xy}$ orbital is slightly larger in the presence of a TiOPc interlayer, suggesting a weakening of the crystal field for VOPc/Ag(100) due to the hybridization with the Ag(100) electrons. Although our fit indicates the $d_{z^2}$ as the highest energy level in all samples, it becomes essentially degenerate with $d_{x^2-y^2}$ for V-O$^{dn}$. As mentioned in the previous section, the relative energy of the $d_{z^2}$ and $d_{x^2-y^2}$ orbitals are critically affected by the changes of the VOPc ligand structure[54, 55]. In the following section, we will delve deeper into this mechanism based on DFT calculations.

## C. Comparison with Density Functional Theory

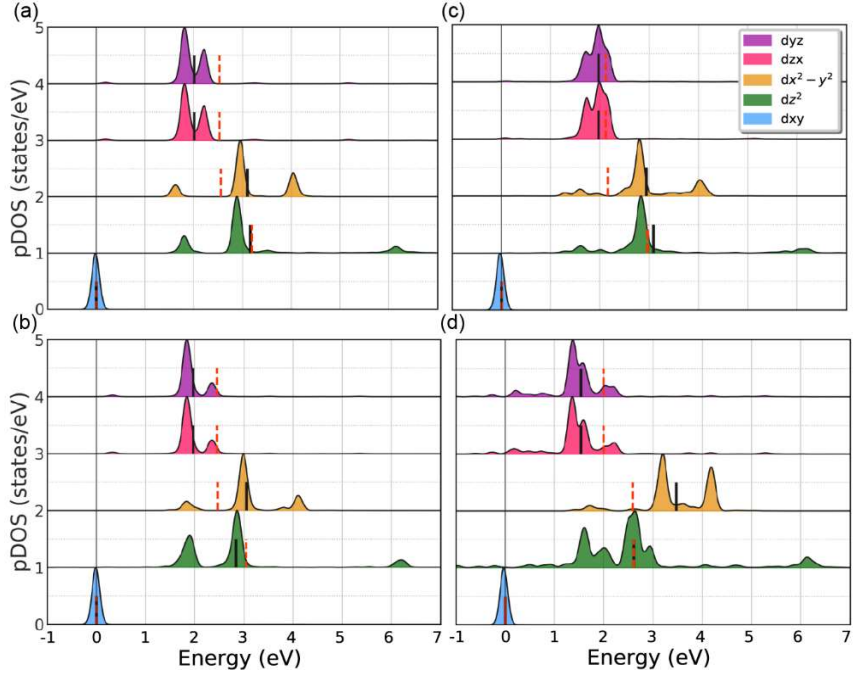

*Figure 6 Projected density of states (pDOS) of d orbitals of (a) VOPc on 2ML TiOPc), (b) VOPc on 1ML TiOPc/Ag(100), (c) oxygen up configuration of VOPc/Ag(100), and (d) oxygen down configuration of VOPc(100) plotted as function of the energy relative to the position of the lowest lying d orbital. The pDOS of the $d_{xy}$ orbital appear as a sharp peak indicating single occupation and atomic like character. The broadening of the other orbitals is due to the hybridization with the ligand and Ag(100) states. Black solid lines show the weighted average energy of each orbital. Red lines are the energy of the 3d orbital found from the average of the best 6 fits to the experiment using multiplet calculations.*

In order to get more insight into the electronic and magnetic properties of the system and further evaluate the validity of our approach, we perform DFT calculations. Differently from the multiplet calculations, this approach incorporates the effect of the hybridization between V, the molecular ligand, and the Ag(100) electrons. Figure 6 presents the projected density of states (pDOS) for the four samples investigated using X-rays and mutiplet calculations, with a focus on the V 3$d$ orbitals. The calculations show a singly occupied $d_{xy}$ with a clear atomic character for all systems. This evidence is in line with our analysis of the XAS data showing a $S$ = 1/2 configuration regardless of the molecular structure. In addition, the atomic-like character of the $d_{xy}$ orbital is not affected by the proximity with the Ag(100), which is a prime requirement to realize robust spin qubits. On the other hand, the unoccupied orbitals show a spread in energy that indicates hybridization with the ligand and/or substrate electrons. In this case, the samples without TiOPc interlayer show a more pronounced energy dispersion, indicating a stronger coupling with the Ag(100) electrons. Moreover, the $d_{zx/zy}$ orbitals exhibit nearly identical pDOS, confirming their degeneracy in line with vanadium's $C_{4v}$ symmetry.

In order to further compare the information derived from DFT with our BO-multiplet approach, we weighted average energy of each orbital pDOS. These orbital energies can be directly compared to the atomic orbital levels obtained from the multiplet calculations. For the majority of orbitals, the multiplet calculations (red dashed line) and DFT (violet solid line) quantitatively agrees (see Fig. 6). While the $d_{z^2}$ energy from the two approaches matches with great accuracy, we notice deviations in the other orbitals, with the multiplet systematically showing lower energies for the $d_{x^2-y^2}$, and higher energies for the $d_{zx/zy}$ orbitals compared to DFT.

These deviations could be traced to the fact that within the DFT model, only the ground state ligand field is modeled while the X-ray spectra are calculated using a point charge representation of the crystal field. Further, the discrepancy could result from a transition between an initial $3d^1$ ground state and a final $2p^5 3d^2$ state, whilst the DFT calculation only accounts for the ground state electronic structure. In the excited state, the presence of the core hole and an additional electron in the valence could alter the crystal field of the molecule compared to the ground state[22]. As the multiplet calculations assumed identical crystal field between initial and final state, the result of the fitting may be seen as an average value between the two configurations, which in turn may differ from the ground state crystal field used in the DFT.

Despite the systematic deviation observed for some orbitals, both models show a good agreement in the trend of energies, with a decreasing tendency to go from higher to lower energies moving from VOPc on 2 ML TiOPc/Ag(100) to V-O$^{dn}$ for both $d_{z^2}$ (Fig. 7a) and $d_{zy/zx}$ (Fig. 7c) which supports our interpretation of a weaker crystal field in the VOPc system in direct contact with the Ag. Furthermore, both models indicate an increase in energy for the $d_{x^2-y^2}$ going from V-O$^{up}$ to V-O$^{dn}$ as shown in Fig. 7b. This effect stems from the larger distortion of the molecule structure for V-O$^{dn}$, where the V atom is closer to the neighboring nitrogen plane with respect to the gas-phase geometry (Fig. 4d). This distortion increases the repulsion between $d_{x^2-y^2}$ and the N atoms, resulting in an increase of energy for this V orbital.

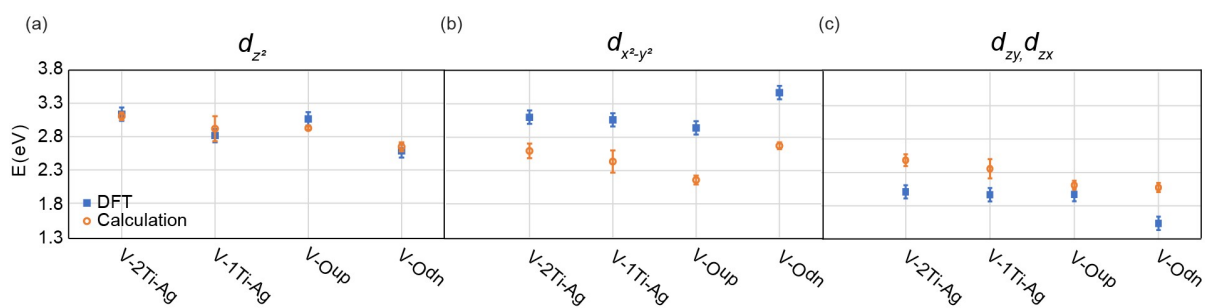

Figure 7 Comparison between energies calculated from DFT and multiplet calculations for the $d_{z^2}$ orbital (a), $d_{x^2-y^2}$ orbital (b) and $d_{zx/zy}$ orbitals (c).

**CONCLUSION**

By combining Bayesian Optimization's uncertainty handling and efficient parameter with atomic multiplet calculations, we developed an efficient tool for X-ray data analysis which can extend the capability of this technique in the investigation of magnetic systems[56]. Specific to the present study, this method allowed us to disentangle the complex influence of many parameters and interpret the fine structure of the experimental XAS spectra at the $L_{2,3}$ edges. Our approach offers a direct way to infer the orbital energy splitting from experiments without additional assumptions or theoretical inputs.

For the VOPc molecular spin qubit candidate, we find them to retain their pristine S=1/2 character even when in direct contact with a metal surface. Conversely, we observe variations in the orbital level splitting depending on the adsorption geometry and molecular film structure, indicating the possibility to tailor their optical properties without altering their spin state. Our results highlight the potential of VOPc as a structurally, electronically, and magnetically robust system for potential implementation in real-world devices.

**ACKNOWLEDGEMENTS**

This work was supported by the Institute for Basic Science (Grant No. IBS-R027-D1). The authors are grateful to W.-D. Schneider for careful reading of the manuscript and valuable comments on this work.

**APPENDIX A: X-ray absorption spectra of VOPc with different molecular film structures**

In Figs. 8-10 we present the comparison between experimental and calculated XAS, XMCD, and XLD for the molecular film structures described in section IIIb, with only the related XMCD at NI summarized in Fig. 4.

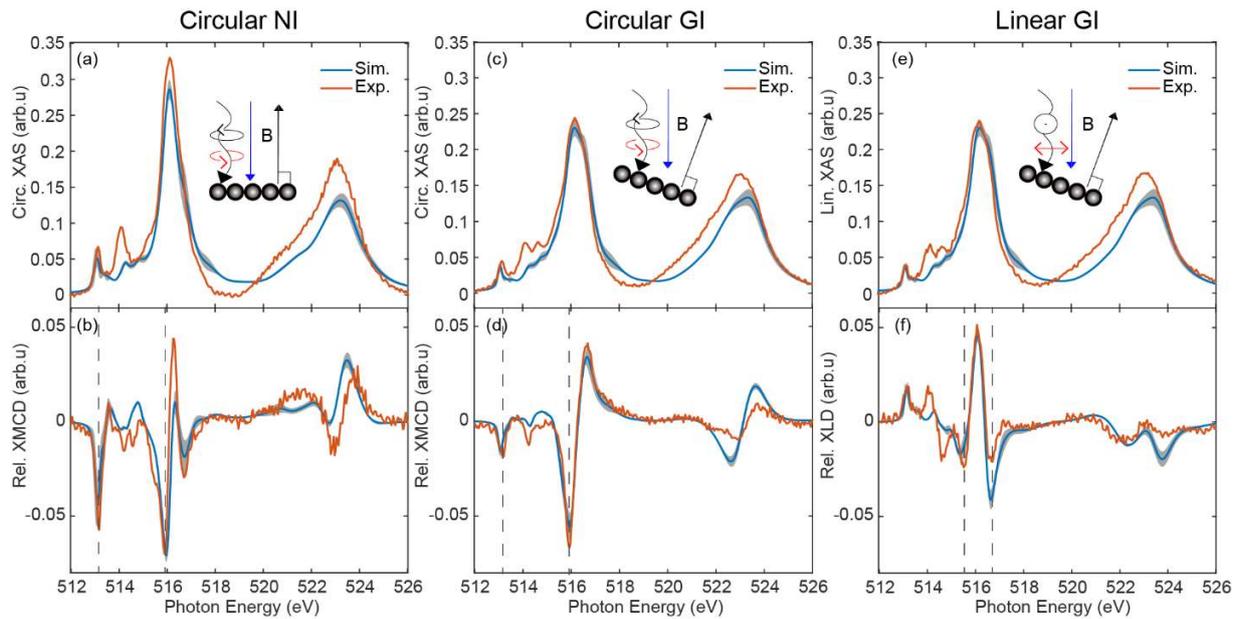

*Figure 8 Experimental and simulated XAS, XMCD, and XLD spectra for VOPc on 1 ML TiOPc/Ag(100). XAS (a) and XMCD (b) spectra at NI and B = 6.8 T, XAS (c) and XMCD (d) spectra at 60° GI and B = 6.8 T, XAS (e) with linear polarization and XLD (f) spectra at GI and B = 0.05 T. The red lines show the experimental data, while blue lines represent the average of the best 6 simulations. The corresponding standard deviation from the mean value at each point is shaded in gray. The features considered to compute the second term in Eq. 4 are indicated by gray dashed lines in (b), (d), and (f).*

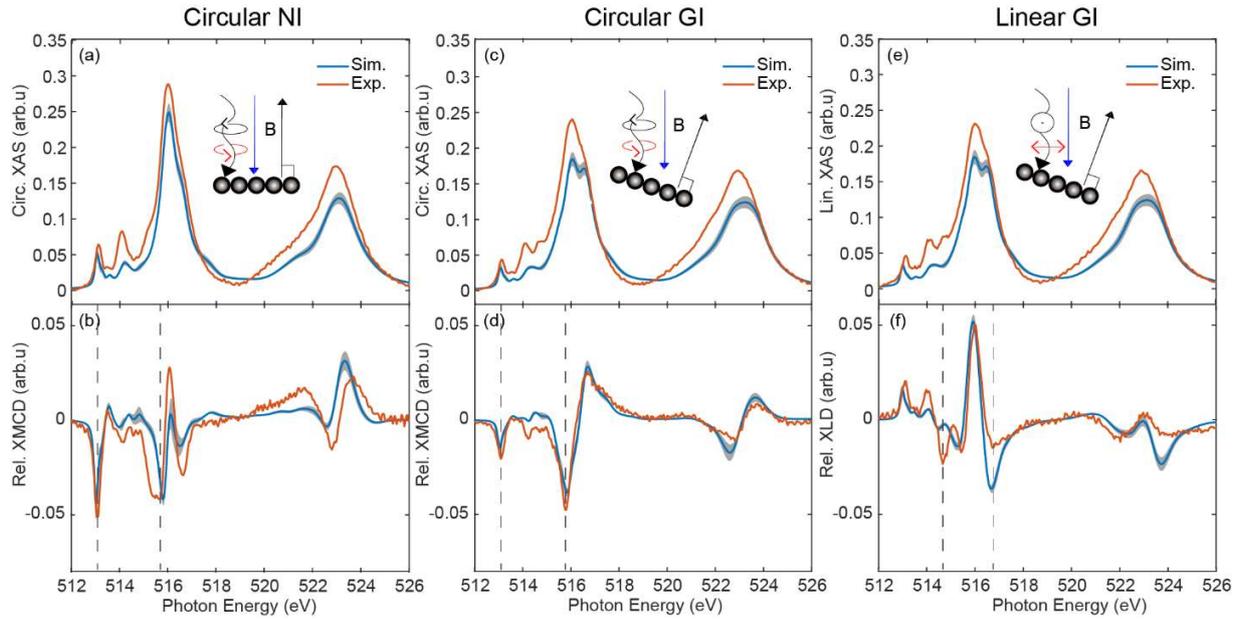

*Figure 9 Experimental and simulated XAS, XMCD, and XLD spectra for VOPc on Ag(100) after annealing. XAS (a) and XMCD (b) spectra at NI and B = 6.8 T, XAS (c) and XMCD (d) spectra at 60° GI and B = 6.8 T, XAS (e) with linear polarization and XLD (f) spectra at GI and B = 0.05 T. The red lines show the experimental data, while blue lines represent the average of the best 6 simulations. The corresponding standard deviation from the mean value at each point is shaded in gray. The features considered to compute the second term in Eq. 4 are indicated by gray dashed lines in (b), (d), and (f).*

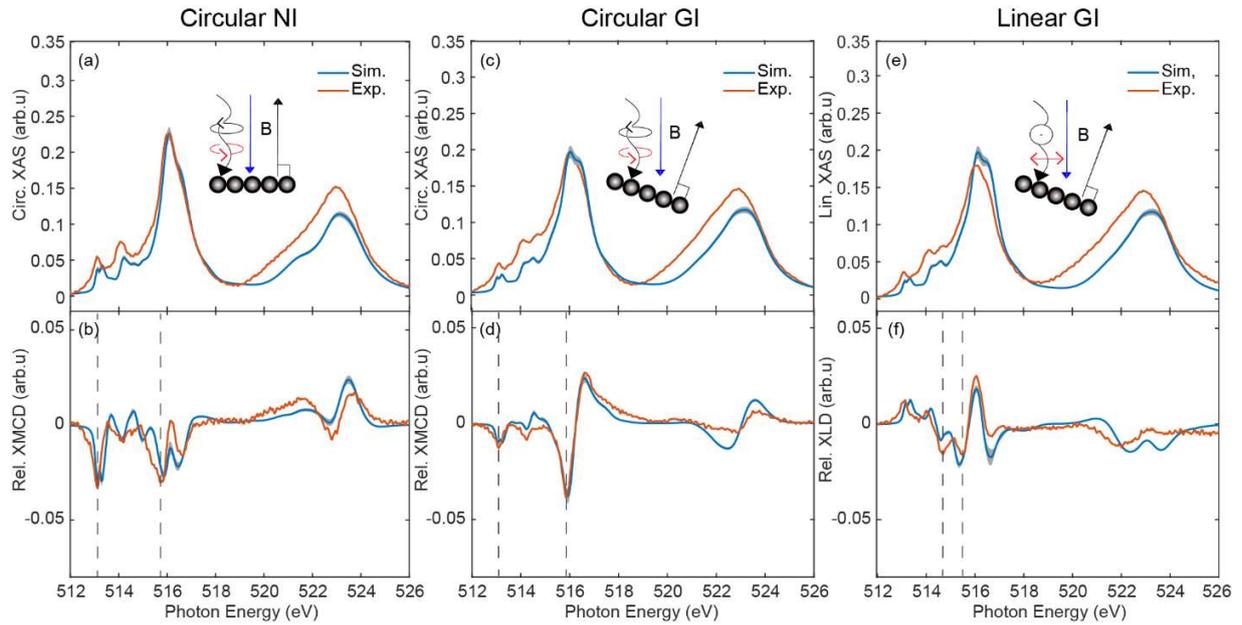

*Figure 10 Experimental and simulated XAS, XMCD, and XLD spectra for VOPc on Ag(100) as deposited. XAS (a) and XMCD (b) spectra at NI and B = 6.8 T, XAS (c) and XMCD (d) spectra at 60° GI and B = 6.8 T, XAS (e) with linear polarization and XLD (f) spectra at GI and B = 0.05 T. The red lines show the experimental data, while blue lines represent the average of the best 6 simulations. The corresponding standard deviation from the mean value at each point is shaded in gray. The features considered to compute the second term in Eq. 4 are indicated by gray dashed lines in (b), (d), and (f).*

**APPENDIX B: Influence of molecular layer structure on the sum rules**

|  | 2 ML of TiOPc on Ag as deposited | 1 ML of TiOPc on Ag as deposited | 0 ML of TiOPc on Ag after annealing | 0 ML of TiOPc on Ag as deposited |
|---|---|---|---|---|
| $\langle S_z \rangle$ | $-0.32\hbar \pm 0.07\,\hbar$ | $-0.41\hbar \pm 0.04\,\hbar$ | $-0.43\hbar \pm 0.05\,\hbar$ | $-0.38\hbar \pm 0.01\,\hbar$ |
| $\langle S_z \rangle_{multiplet}$ | $-0.49\,\hbar$ | $-0.49\,\hbar$ | $-0.49\,\hbar$ | $-0.49\,\hbar$ |
| $\langle L_z \rangle$ | $-0.05 \pm 0.01\,\hbar$ | $-0.23 \pm 0.02\,\hbar$ | $-0.35 \pm 0.03\,\hbar$ | $-0.08 \pm 0.01\,\hbar$ |
| $\langle L_z \rangle_{multiplet}$ | $0.01\,\hbar$ | $0.01\,\hbar$ | $0.02\,\hbar$ | $0.01\,\hbar$ |

*Table 2 Comparison between spin and orbital angular momentum calculated using sum rules and multiplet calculations.*

Spin and orbital angular momenta obtained from sum rules are summarized in Table 2. As discussed in the main text, the values of the spin are lower than what expected for a S = 1/2 system due the overlap between the $L_3$ and $L_2$ edges[53]. On the other hand, the multiplet calculations find the expected values for all systems. For the orbital angular momentum, sum rules always find values with the same sign of the spin, which is not expected for less than half filled shell system, also in disagreement with our multiplet calculations. The inconsistency of the sign of the orbital angular momentum sum rules could be due to an asymmetric matrix element for left and right polarizations affecting the dipole transitions, as previously observed at the $L_{2,3}$ edges of lanthanides [57]. In the present case, this asymmetry may originate from a spin-dependent hybridization of the V 3$d$ orbitals with the molecular ligand and may prevent the use of sum rules to effectively quantify the angular momenta of the system.

**APPENDIX C: Rescaling factors for the Slater integrals used in the multiplet calculations**

|  | V-2Ti-Ag | V-1Ti-Ag | V-O[up] | V-O[dn] |
|---|---|---|---|---|
| Res. 2$p$-3$d$ | 0.51 ± 0.02 | 0.53 ± 0.03 | 0.56 ± 0.02 | 0.45 ± 0.01 |
| Res. 3$d$-3$d$ | 0.63 ± 0.02 | 0.58 ± 0.02 | 0.52 ± 0.03 | 0.58 ± 0.02 |

*Table 3 Rescaling factors obtained from the BO and used in the multiplet calculations. The rescaling factors are applied to the atomic values obtained from the Cowan code.[41]*

The values of the Slater integrals and spin-orbit coupling were determined by using atomic calculations from the Cowan code[41] and rescale them by the factors indicated in Table 3. The rescaling factors for the atomic values of the Slater integrals obtained in this study are significantly lower than the typical values used for other transition metals[22]. This indicates that there is a significant hybridization between the atomic states and the ligand, leading to a reduction of the intraatomic correlations. Our fit yields the same rescaling factors for V-2Ti-Ag and V-1Ti-Ag, indicating a higher degree of decoupling from the substrate. On the other hand, the fit show a slightly lower values for V-O[up] and V-O[dn], suggesting a higher level of hybridization with the silver substrate. This finding suggests that there is a non-negligible influence of the metal surface on the electronic state of V. This may also indicate a limitation of the crystal field model used in the multiplet calculations, which requires an unusually low rescaling of the atomic values of the Slater integrals to match the experimental line shape.

**APPENDIX D: Combined BO-multiplet calculations using only residual sum of the squares**

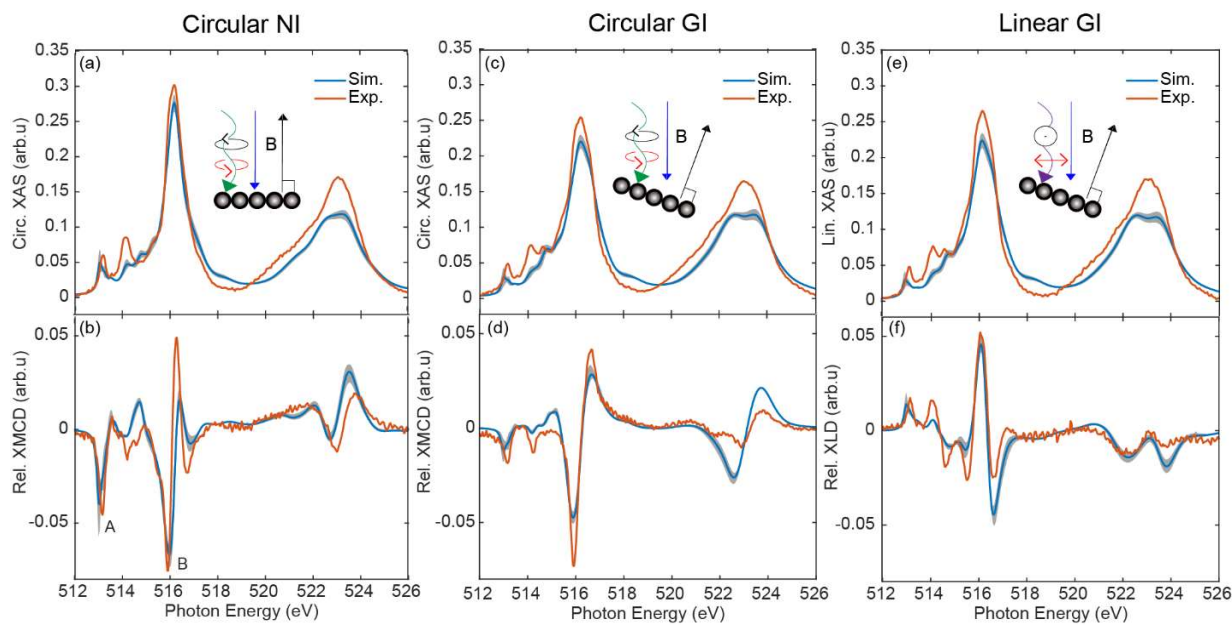

*Figure 11: Experimental and simulated XAS, XMCD, and XLD spectra for VOPc on 2 ML of TiOPc/Ag (100). XAS (a) and XMCD (b) spectra at NI and B = 6.8 T, XAS (c) and XMCD (d) spectra at 60° GI and B = 6.8 T, XAS (e) with linear polarization and XLD (f) spectra at GI and B = 0.05 T. The red lines show the experimental data, while blue lines represent the average of the best 6 simulations. The corresponding standard deviation from the mean value at each point is shaded in gray.*

In Figure 11, we present the results of combined BO-multiplet calculations without the additional contribution of the relevant features in the objective function ($W$=0 in Eq. 4). While these results demonstrate a good fit, the sharp features in the $L_3$ XMCD indicated as A and B are not accurately reproduced. The method also fails to capture the trend of these features of the XMCD spectra at NI over the set of samples investigated, as shown in Fig. 12. The lack of accuracy in capturing these features supports the use of a weighting term (W > 0 in Eq. 4) to correctly fit the data and infer the related orbital structure.

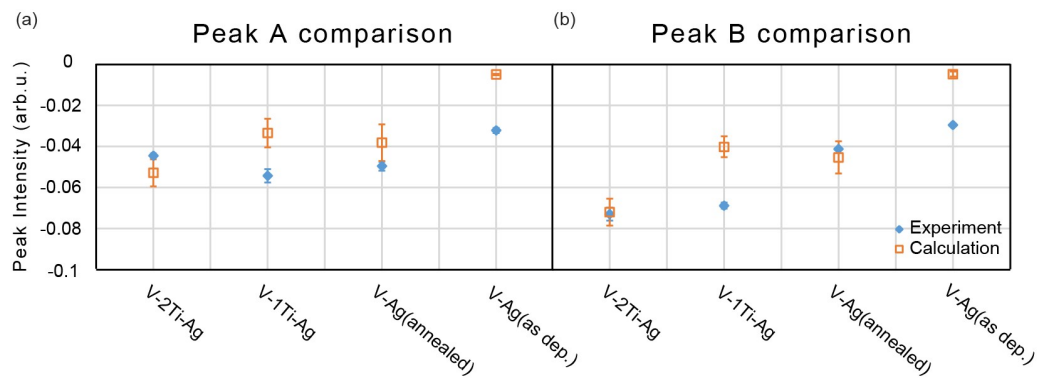

*Figure 12: Comparison between experiment and multiplet calculations with (W=0 in Eq. 4) for the intensity of the peak A (a) and peak B (b) of the XMCD at NI over the different samples. For the experiments, we considered the average of the three points at the tip of each peak and the error bars show the standard deviation over these three points. For multiplet calculations, we considered the average of the tipping points of each peak over 6 calculations and the error bars show the standard deviation over these six values.*

**APPENDIX E: Selected-windows method in combined BO-multiplet calculations**

In this section we present an alternative approach to fit the total XAS, XMCD, and XLD of VOPc deposited on 2ML of TiOPc/Ag(100). Differently from Eq. 4, where we used a discrete set of single points weighted in the error function, here we select spectral regions included in energy windows around the features of interest through the error function:

$$Err_{window} = \sum_\alpha \frac{RSS_\alpha}{TSS_\alpha} + W \sum_\alpha \sum_i (\sigma_{\alpha,i} - \hat{\sigma}_{\alpha,i})^2 \qquad \text{(Eq. 5)}$$

In Eq. 5, the first term is ratio between the residual sum of squares $RSS_\alpha$ computed for each experimental/simulated spectra pair and the total sum of squares ($TSS_\alpha$) of the experimental spectra, with $\alpha$ labeling each of the 6 XAS, XMCD, and XLD. They are calculated in the same way as Eq. 4. The second term represents the selected windows, which include 11 points (corresponded to a window of 0.4 eV) centered around the most relevant features from the experiments ($\sigma_{\alpha,i}$). This term in the error function is calculated by squaring the residuals with respect to the calculated spectrum in the same energy range ($\hat{\sigma}_{\alpha,i}$), where *i* labels the windows around two prominent features in each XMCD and XLD spectra, for a total of 6 windows for each sample. Similar to Eq.4, we introduced a coefficient *W* to equilibrate the significance of the terms, ensuring their comparable relevance in the optimization procedure and enhancing the accuracy of the fit of the relevant features.

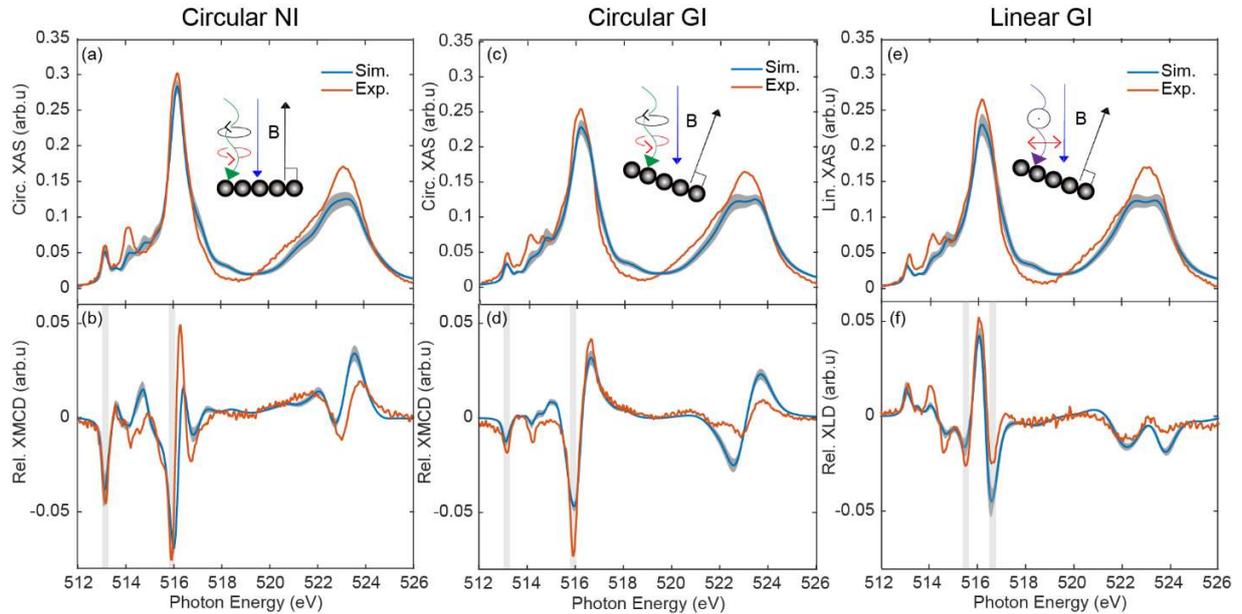

*Figure 13: Experimental and simulated XAS, XMCD, and XLD spectra for VOPc on 2 ML of TiOPc/Ag (100). XAS (a) and XMCD (b) spectra at NI and B = 6.8 T, XAS (c) and XMCD (d) spectra at 60° GI and B = 6.8 T, XAS (e) with linear polarization and XLD (f) spectra at GI and B = 0.05 T. The red lines show the experimental data, while blue lines represent the average of the best 6*

simulations. The corresponding standard deviation from the mean value at each point is shaded in gray around the blue line. The selected windows used to compute the second term in Eq. 5 are filled in gray in (b), (d) and (f).

For the method employed, we also conducted tests to assign appropriate weights, aiming to balance the influence of the selected windows and ensure their comparable significance in the optimization process. The data presented in Fig. 13 were derived assigning a weight $W$ = 1 to the spectral features within the marked gray windows. This approach revealed a minor discrepancy in the XAS data; however, the overall fit is notably satisfactory with only very minor differences observable with respect to the fit performed using Eq. 4 as the error function.

| Res. 2p-3d | Res. 3d-3d | $B_{1g}$ ($d_{x^2-y^2}$) | $A_{1g}$ ($d_{z^2}$) | $E_g$ ($d_{xz,yz}$) | Ampl. |
|---|---|---|---|---|---|
| 0.51 ± 0.01 | 0.70 ± 0.04 | 2.58 ± 0.07 | 3.23 ± 0.10 | 2.60 ± 0.07 | 0.95 ± 0.08 |

*Table 4 Parameters obtained from the windows selected method in combined BO-multiplet calculations, with W = 1.*

The values of the free parameters, obtained through the windows selected method in combined BO-multiplet calculations, are indicated in Table 4. Regarding the rescaling factor of the Slater integrals, they are slightly larger than the values shown in Table 3. For the crystal field levels, the only difference is found on the $d_{zy/zx}$ orbital, which is now found at 0.1 eV higher than the corresponding value show in Table 1.

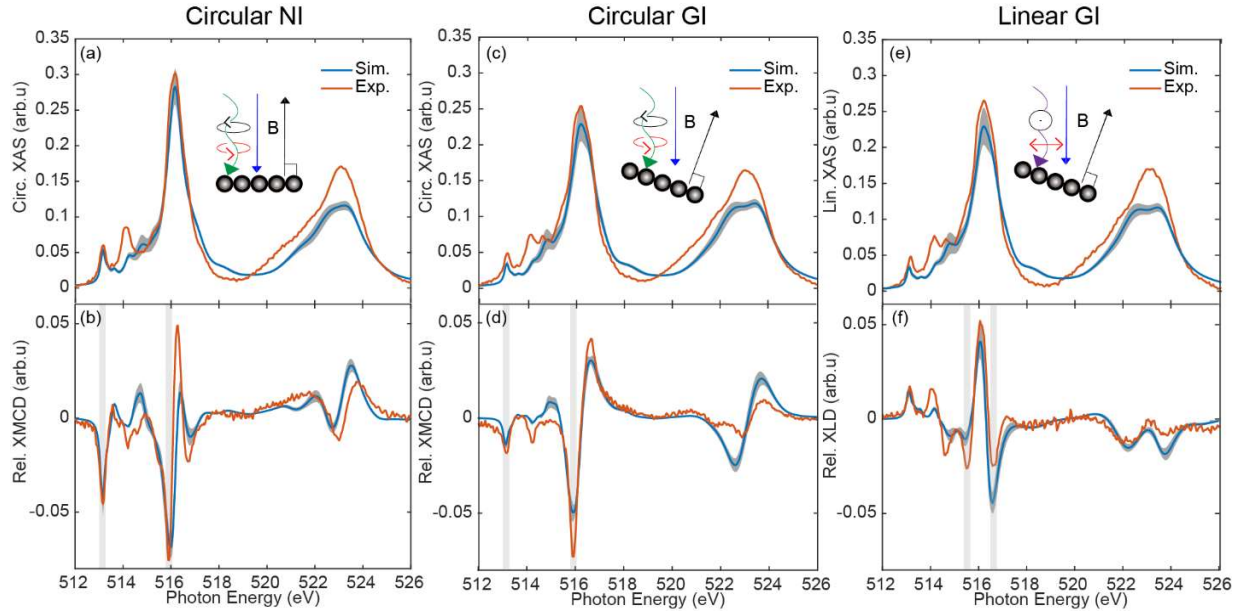

*Figure 14: Experimental and simulated XAS, XMCD, and XLD spectra for VOPc on 2 ML of TiOPc/Ag (100). XAS (a) and XMCD (b) spectra at NI and B = 6.8 T, XAS (c) and XMCD (d) spectra at 60° GI and B = 6.8 T, XAS (e) with linear polarization and XLD (f) spectra at GI and B = 0.05 T. The red lines show the experimental data, while blue lines represent the average of the best 6 simulations. The corresponding standard deviation from the mean value at each point is shaded in gray around the blue line. The selected windows used to compute the second term in Eq. 5 are filled in gray in (b), (d) and (f).*

The data presented in Fig. 14 were derived using a weighting factor $W$ = 30. The results of this fitting procedure are very similar to those shown in Fig. 13.

| Res. 2p-3d | Res. 3d-3d | $B_{1g}$ ($d_{x^2-y^2}$) | $A_{1g}$ ($d_{z^2}$) | $E_g$ ($d_{xz,yz}$) | Ampl. |
|---|---|---|---|---|---|
| 0.51 ± 0.02 | 0.68 ± 0.03 | 2.49 ± 0.06 | 3.10 ± 0.05 | 2.58 ± 0.06 | 0.88 ± 0.05 |

*Table 5 Parameters obtained from the windows selected method in combined BO-multiplet calculations, with W = 30.*

The values of the free parameters obtained through this method shown in Table 5 are very similar to those of Table 4 obtained with W = 1, with crystal field parameters differing only by about 0.1 eV at most. Therefore, it can be concluded that larger weighting to the selected windows is unnecessary to improve the fit quality.

The selected-window approach also allows us to achieve a good match with the data, however, without improving the accuracy of the fit. For instance, a minor energy mismatch is still visible on the second feature of the NI XMCD, and it could possibly be attributed to inherent limitations of the multiplet calculations. In addition, the choice of the window range represents an additional arbitrarily chosen parameter. Therefore, in this study, we opted to employ the single-point-per-feature descriptor shown in Eq. 4 as it offers the best trade-off between required fit accuracy and the number of arbitrarily chosen parameters.

**APPENDIX F: Partial dependence plot of combined BO-multiplet calculations**

Partial Dependence Plots (PDPs) are a powerful visualization tool that aids in the interpretation of machine learning models and helps in understanding the relationships between the target variable and individual features[58]. They offer valuable insights into whether these relationships are linear, monotonic, or exhibit more intricate behaviors. Partial Dependence plots (obtained from the MATLAB function *partialDependence*) show the predicted responses on a subset of predictor variables by marginalizing over the other variables[59]. Hence, they offer the possibility to represent the distribution obtained from a Gaussian process regression for multi-dimensional parameter space. In the case of our study, we use n = 6 independent fit parameters, hence this representation requires $\frac{n(n-1)}{2} = 15$ independent plots.

In Figures 15 to 18, we present the PDPs for the combined BO-multiplet fits to the data of the four samples presented in Fig 4. The plots are related to one of the best 6 fits performed for each sample and illustrate the correlation between various parameters. Each plot represents the parameter region of the two indicated variables where the color scale corresponds to the error function marginalized over the other variables. It is clearly visible that each PDP shows a minimum in the considered range, indicating a correct choice of the parameters space.

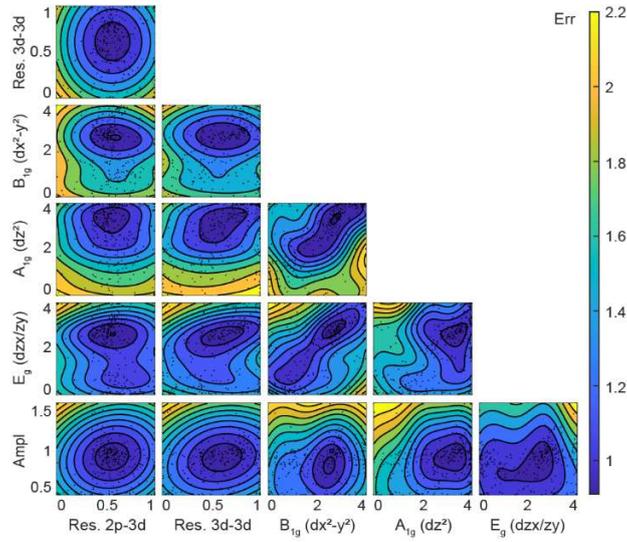

*Figure 15: Partial dependence plot of the BO used for VOPc on 2 ML of TiOPc/Ag (100). Each plot shows the dependence of the error function over the two considered variables, after marginalization over the other variables. The color scale represents the value of the error function after marginalization. Dots on the plots mark the observation on the subspace of the indicated variables.*

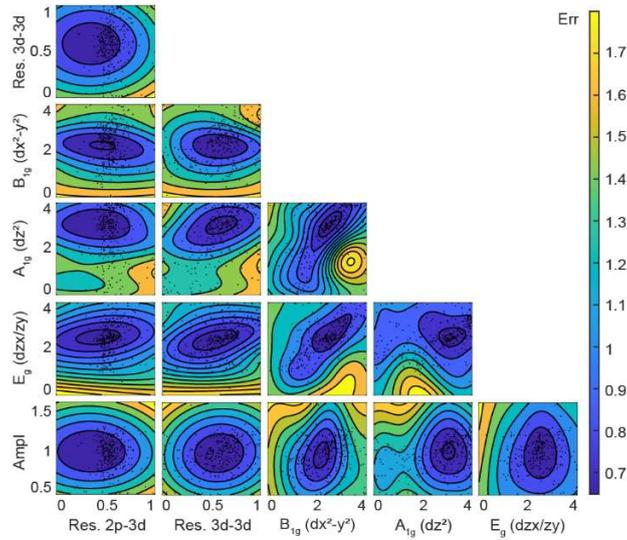

*Figure 16: Partial dependence plot of the BO used for VOPc on 1 ML of TiOPc/Ag (100). Each plot shows the dependence of the error function over the two considered variables, after marginalization over the other variables. The color scale represents the value of the error function after marginalization. Dots on the plots mark the observation on the subspace of the indicated variables.*

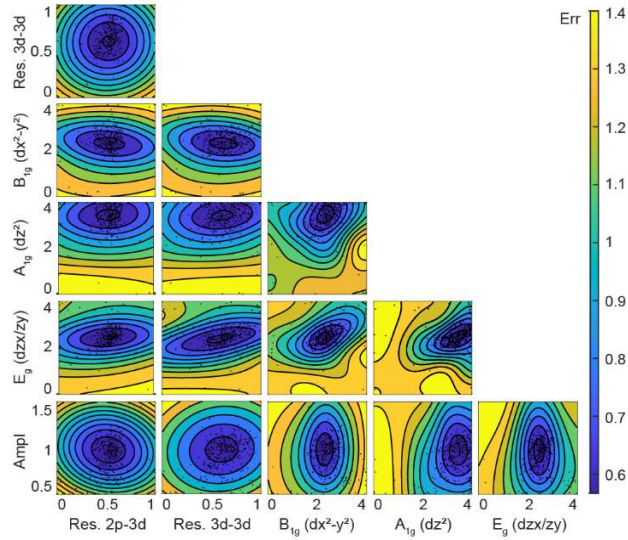

*Figure 17: Partial dependence plot of the BO used for VOPc on Ag (100), after annealing. Each plot shows the dependence of the error function over the two considered variables, after marginalization over the other variables. The color scale represents the value of the error function after marginalization. Dots on the plots mark the observation on the subspace of the indicated variables.*

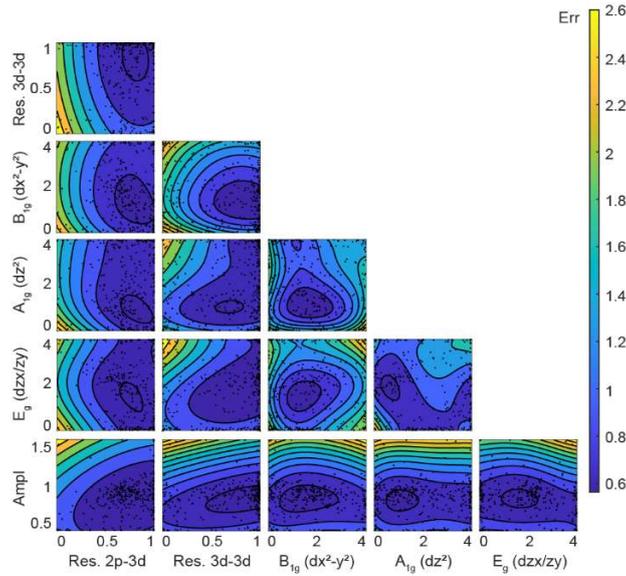

*Figure 18: Partial dependence plot of the BO used for VOPc on Ag (100), as deposited. Each plot shows the dependence of the error function over the two considered variables, after marginalization over the other variables. The color scale represents the value of the error function after marginalization. Dots on the plots mark the observation on the subspace of the indicated variables.*

1.  Nielsen, M.A. and I.L. Chuang, *Quantum computation and quantum information*. 2010: Cambridge university press.
2.  Gaita-Ariño, A., et al., *Molecular spins for quantum computation.* Nature chemistry, 2019. **11**(4): p. 301-309.
3.  Atzori, M. and R. Sessoli, *The second quantum revolution: role and challenges of molecular chemistry.* Journal of the American Chemical Society, 2019. **141**(29): p. 11339-11352.
4.  Graham, M.J., et al., *Forging solid-state qubit design principles in a molecular furnace.* Chemistry of Materials, 2017. **29**(5): p. 1885-1897.
5.  Coronado, E., *Molecular magnetism: from chemical design to spin control in molecules, materials and devices.* Nature Reviews Materials, 2020. **5**(2): p. 87-104.
6.  Wruss, E., et al., *Magnetic configurations of open-shell molecules on metals: The case of CuPc and CoPc on silver.* Physical Review Materials, 2019. **3**(8): p. 086002.
7.  Zhang, X., et al., *Electron spin resonance of single iron phthalocyanine molecules and role of their non-localized spins in magnetic interactions.* Nature Chemistry, 2022. **14**(1): p. 59-65.
8.  Adler, H., et al., *Interface properties of VOPc on Ni (111) and graphene/Ni (111): Orientation-dependent charge transfer.* The Journal of Physical Chemistry C, 2015. **119**(16): p. 8755-8762.
9.  Eguchi, K., et al., *Direct synthesis of vanadium phthalocyanine and its electronic and magnetic states in monolayers and multilayers on Ag (111).* The Journal of Physical Chemistry C, 2015. **119**(18): p. 9805-9815.
10. Warner, M., et al., *Potential for spin-based information processing in a thin-film molecular semiconductor.* Nature, 2013. **503**(7477): p. 504-508.
11. Atzori, M., et al., *Room-temperature quantum coherence and rabi oscillations in vanadyl phthalocyanine: toward multifunctional molecular spin qubits.* Journal of the American Chemical Society, 2016. **138**(7): p. 2154-2157.
12. Blowey, P.J., et al., *The Structure of VOPc on Cu (111): Does V═O Point Up, or Down, or Both?* The Journal of Physical Chemistry C, 2018. **123**(13): p. 8101-8111.
13. Eguchi, K., et al., *Molecular orientation and electronic states of vanadyl phthalocyanine on si (111) and ag (111) surfaces.* The Journal of Physical Chemistry C, 2013. **117**(44): p. 22843-22851.
14. Eguchi, K., et al., *Magnetic interactions of vanadyl phthalocyanine with ferromagnetic iron, cobalt, and nickel surfaces.* The Journal of Physical Chemistry C, 2014. **118**(31): p. 17633-17637.
15. Malavolti, L., et al., *Tunable spin–superconductor coupling of spin 1/2 vanadyl phthalocyanine molecules.* Nano Letters, 2018. **18**(12): p. 7955-7961.
16. Cimatti, I., et al., *Vanadyl phthalocyanines on graphene/SiC (0001): toward a hybrid architecture for molecular spin qubits.* Nanoscale Horizons, 2019. **4**(5): p. 1202-1210.
17. Noh, K., et al., *Template-directed 2D nanopatterning of S= 1/2 molecular spins.* Nanoscale Horizons, 2023. **8**(5): p. 624-631.
18. Nanba, Y., et al., *Configuration-interaction full-multiplet calculation to analyze the electronic structure of a cyano-bridged coordination polymer electrode.* The Journal of Physical Chemistry C, 2012. **116**(47): p. 24896-24901.
19. Konecny, L., et al., *Accurate x-ray absorption spectra near L-and M-edges from relativistic four-component damped response time-dependent density functional theory.* Inorganic Chemistry, 2021. **61**(2): p. 830-846.
20. Besley, N.A., *Density functional theory based methods for the calculation of x-ray spectroscopy.* Accounts of Chemical Research, 2020. **53**(7): p. 1306-1315.
21. Josefsson, I., et al., *Ab initio calculations of x-ray spectra: Atomic multiplet and molecular orbital effects in a multiconfigurational scf approach to the l-edge spectra of transition metal complexes.* The journal of physical chemistry letters, 2012. **3**(23): p. 3565-3570.

22. De Groot, F., *Multiplet effects in X-ray spectroscopy.* Coordination Chemistry Reviews, 2005. **249**(1-2): p. 31-63.
23. Gallardo, I., et al., *Large effect of metal substrate on magnetic anisotropy of Co on hexagonal boron nitride.* New Journal of Physics, 2019. **21**(7): p. 073053.
24. Cardot, C., et al., *Core-to-Core X-ray Emission Spectra from Wannier Based Multiplet Ligand Field Theory.* arXiv preprint arXiv:2304.14582, 2023.
25. Herrera-Yáñez, M.G., et al., *Fitting Multiplet Simulations to L-Edge XAS Spectra of Transition-Metal Complexes Using an Adaptive Grid Algorithm.* Inorganic Chemistry, 2023. **62**(9): p. 3738-3760.
26. Chen, A., X. Zhang, and Z. Zhou, *Machine learning: accelerating materials development for energy storage and conversion.* InfoMat, 2020. **2**(3): p. 553-576.
27. Torun, H.M., et al., *A global Bayesian optimization algorithm and its application to integrated system design.* IEEE Transactions on Very Large Scale Integration (VLSI) Systems, 2018. **26**(4): p. 792-802.
28. Moriconi, R., M.P. Deisenroth, and K. Sesh Kumar, *High-dimensional Bayesian optimization using low-dimensional feature spaces.* Machine Learning, 2020. **109**: p. 1925-1943.
29. Kanazawa, T., *Efficient Bayesian Optimization using Multiscale Graph Correlation.* arXiv preprint arXiv:2103.09434, 2021.
30. Carbone, M.R., et al., *Machine-learning X-ray absorption spectra to quantitative accuracy.* Physical review letters, 2020. **124**(15): p. 156401.
31. Zhang, Y., et al., *Autonomous atomic Hamiltonian construction and active sampling of X-ray absorption spectroscopy by adversarial Bayesian optimization.* npj Computational Materials, 2023. **9**(1): p. 46.
32. Carbone, M.R., et al., *Classification of local chemical environments from x-ray absorption spectra using supervised machine learning.* Physical Review Materials, 2019. **3**(3): p. 033604.
33. Li, L., M. Lu, and M.K. Chan, *A Deep Learning Model for Atomic Structures Prediction Using X-ray Absorption Spectroscopic Data.* arXiv preprint arXiv:1905.03928, 2019.
34. Rankine, C.D., M.M. Madkhali, and T.J. Penfold, *A deep neural network for the rapid prediction of X-ray absorption spectra.* The Journal of Physical Chemistry A, 2020. **124**(21): p. 4263-4270.
35. Ghosh, K., et al., *Deep learning spectroscopy: Neural networks for molecular excitation spectra.* Advanced science, 2019. **6**(9): p. 1801367.
36. Carra, P., et al., *X-ray circular dichroism and local magnetic fields.* Physical Review Letters, 1993. **70**(5): p. 694.
37. Piamonteze, C., et al., *X-Treme beamline at SLS: X-ray magnetic circular and linear dichroism at high field and low temperature.* Journal of synchrotron radiation, 2012. **19**(5): p. 661-674.
38. Stöhr, J., *X-ray magnetic circular dichroism spectroscopy of transition metal thin films.* Journal of Electron Spectroscopy and Related Phenomena, 1995. **75**: p. 253-272.
39. Thole, B., et al., *X-ray circular dichroism as a probe of orbital magnetization.* Physical review letters, 1992. **68**(12): p. 1943.
40. Haverkort, M.W. *Quanty for core level spectroscopy-excitons, resonances and band excitations in time and frequency domain.* in *Journal of Physics: Conference Series.* 2016. IOP Publishing.
41. Cowan, R.D., *The theory of atomic structure and spectra.* 1981: Univ of California Press.
42. Krause, M.O. and J. Oliver, *Natural widths of atomic K and L levels, K α X-ray lines and several KLL Auger lines.* Journal of Physical and Chemical Reference Data, 1979. **8**(2): p. 329-338.
43. Frazier, P.I., *A tutorial on Bayesian optimization.* arXiv preprint arXiv:1807.02811, 2018.
44. Guda, A.A., et al., *Understanding X-ray absorption spectra by means of descriptors and machine learning algorithms.* npj Computational Materials, 2021. **7**(1): p. 203.

45. Giannozzi, P., et al., *QUANTUM ESPRESSO: a modular and open-source software project for quantum simulations of materials.* Journal of physics: Condensed matter, 2009. **21**(39): p. 395502.
46. Dal Corso, A., *Pseudopotentials periodic table: From H to Pu.* Computational Materials Science, 2014. **95**: p. 337-350.
47. Perdew, J.P., K. Burke, and M. Ernzerhof, *Generalized gradient approximation made simple.* Physical review letters, 1996. **77**(18): p. 3865.
48. Sabatini, R., T. Gorni, and S. De Gironcoli, *Nonlocal van der Waals density functional made simple and efficient.* Physical Review B, 2013. **87**(4): p. 041108.
49. Colonna, S., et al., *Supramolecular and Chiral Effects at the Titanyl Phthalocyanine/Ag (100) Hybrid Interface.* The Journal of Physical Chemistry C, 2014. **118**(10): p. 5255-5267.
50. Xu, Z., et al., *Orienting dilute thin films of non-planar spin-1/2 vanadyl–phthalocyanine complexes.* Materials Advances, 2022. **3**(12): p. 4938-4946.
51. Assour, J., J. Goldmacher, and S. Harrison, *Electron spin resonance of vanadyl phthalocyanine.* The Journal of Chemical Physics, 1965. **43**(1): p. 159-165.
52. Tverdova, N.V., et al., *The molecular structure, bonding, and energetics of oxovanadium phthalocyanine: an experimental and computational study.* Structural Chemistry, 2013. **24**: p. 883-890.
53. Piamonteze, C., P. Miedema, and F.M. De Groot, *Accuracy of the spin sum rule in XMCD for the transition-metal L edges from manganese to copper.* Physical Review B, 2009. **80**(18): p. 184410.
54. Ziolo, R.F., C.H. Griffiths, and J.M. Troup, *Crystal structure of vanadyl phthalocyanine, phase II.* Journal of the Chemical Society, Dalton Transactions, 1980(11): p. 2300-2302.
55. Cranston, R.R. and B.H. Lessard, *Metal phthalocyanines: Thin-film formation, microstructure, and physical properties.* RSC advances, 2021. **11**(35): p. 21716-21737.
56. Vaňo, V., et al., *Emergence of Exotic Spin Texture in Supramolecular Metal Complexes on a 2D Superconductor.* arXiv preprint arXiv:2309.02537, 2023.
57. van Veenendaal, M., J.B. Goedkoop, and B.T. Thole, *Branching Ratios of the Circular Dichroism at Rare Earth $L_{23}$ Edges.* Physical Review Letters, 1997. **78**(6): p. 1162-1165.
58. Moosbauer, J., et al., *Explaining hyperparameter optimization via partial dependence plots.* Advances in Neural Information Processing Systems, 2021. **34**: p. 2280-2291.
59. Friedman, J.H., *Greedy function approximation: a gradient boosting machine.* Annals of statistics, 2001: p. 1189-1232.